# A HIGH-PERFORMANCE COBALT-FREE CATHODE FOR PROTON-CONDUCTING SOLID OXIDE FUEL CELLS VIA MULTI-ELEMENT DOPING IN $Sr_2Fe_2O_6$

Le Zhou [a], Yanru Yin [a], Dilshod Nematov [b], Hailu Dai [c], Yuyuan Gu [a], Shoufu Yu [a,*] and Lei Bi [a,*]

[a] Fuel cell group, School of Resource Environment and Safety Engineering, University of South China, Hengyang 421001, China

[b] S.U. Umarov Physical-Technical Institute of NAST, Dushanbe 734063, Tajikistan

[c] School of Materials Science and Engineering, Yancheng Institute of Technology, Yancheng 224051, China

*Corresponding authors: YSF20230303@163.com (S.F. Yu) and Lei.Bi@usc.edu.cn; bilei81@gmail.com

**Abstract.** The development of efficient and stable intermediate-temperature solid oxide fuel cells (SOFCs) necessitates high-performance cathode materials that are cobalt-free, cost-effective, and compatible with proton-conducting electrolytes. While $Sr_2Fe_2O_6$ (SFO)-based ferrites offer a promising cobalt-free alternative, their electrochemical performance requires further enhancement to compete with state-of-the-art cathodes. This study proposes and validates a multi-element doping strategy as a superior approach to tailor the properties of SFO. The specific oxide $Sr_2Fe_{1.5}Mo_{0.125}Sn_{0.125}Sc_{0.125}Zr_{0.125}O_6$ (SFO-ZSSM) is designed, synthesized via a solid-state reaction method, and systematically evaluated as a cathode for proton-conducting SOFCs (H-SOFCs). Its performance is benchmarked against a series of SFO cathodes modified with single dopants (Mo, Sn, Sc, Zr). Structural characterization confirms the successful formation of a phase-pure perovskite structure with homogeneous elemental distribution. Electrical conductivity relaxation (ECR) measurements reveal that SFO-ZSSM exhibits dramatically enhanced oxygen and proton transport kinetics compared to all singly-doped counterparts, demonstrating a significant synergistic effect. Consequently, fuel cells employing the SFO-ZSSM cathode deliver exceptional peak power densities of 1580, 1137, and 854 mW cm$^{-2}$ at 700, 650, and 600 °C, respectively, significantly outperforming cells with single-doped cathodes. Electrochemical impedance spectroscopy further corroborates its superior catalytic activity, showing the lowest polarization resistance. Moreover, the SFO-ZSSM cell demonstrates excellent operational stability over 100 hours, attributed to its robust microstructure and Ba-free composition. This work conclusively establishes the multi-element doping strategy as a highly effective pathway for engineering high-performance, cobalt-free cathodes for next-generation H-SOFCs.

***Keywords*:** Multi-element doping; Cathode; Proton conductor; SOFCs

# 1. Introduction

Sustainable materials and technologies present promising solutions to contemporary energy and environmental challenges, garnering considerable attention[1-4]. Fuel cells, which directly convert the chemical energy of fuels into electricity, are regarded as highly efficient devices for electrical power generation[5-7]. Solid oxide fuel cells (SOFCs), a major branch of fuel cell technology, inherit the characteristic advantages of high efficiency and low pollution while also possessing unique benefits such as all-solid-state structures and considerable flexibility in fuel selection[8]. Furthermore, SOFCs enable the utilization of metal oxides instead of precious noble metals as electrode catalysts, which contributes to significant cost reduction[9, 10]. However, conventional SOFCs typically operate at elevated temperatures (above 800°C), leading to several inherent problems including material degradation, sealing difficulties, and long startup times[11]. Consequently, reducing the operational temperature of SOFCs has become a primary objective for advancing this technology[12].

SOFCs employing proton-conducting electrolytes, referred to as proton-conducting SOFCs (H-SOFCs)[13] or protonic ceramic fuel cells (PCFCs)[14], offer a viable pathway to lower the working temperature. Due to the high ionic conductivity and low activation energy associated with proton-conducting electrolytes[15, 16], the resistance contribution from the electrolyte remains relatively small even at intermediate temperatures[17, 18]. However, lowering the operating temperature also results in sluggish cathode kinetics, causing the cathode polarization resistance to become the dominant performance-limiting factor[19-21]. Therefore, the development of suitable high-performance cathode materials specifically for H-SOFCs is essential[22, 23]. Over the past decade, numerous cathode materials have been proposed[24-27], and several have indeed demonstrated promising performance for H-SOFC applications[28, 29]. It is noteworthy, however, that most of these high-performing cathodes utilize cobalt as a primary constituent[30-33]. Although cobalt-containing cathodes generally offer high catalytic activity, they suffer from several drawbacks such as high thermal expansion coefficients, potential cobalt evaporation, and the high cost associated with cobalt[34, 35]. This has driven research towards designing cobalt-free cathodes for H-SOFCs, making it a prominent topic in the field[36] and leading to the proposal of ferrite-based cathodes due to their cobalt-free composition and acceptable initial performance[37, 38]. Among various ferrite-based cathodes, $Sr_2Fe_2O_6$ (SFO)-based materials have recently attracted considerable attention. While traditional, unmodified SFO exhibits inadequate performance for H-SOFCs, the strategic tailoring of SFO with selected dopants has successfully improved its functionality[39]. Nevertheless, the performance enhancement achieved with currently modified SFO cathodes still cannot compete with the best-performing cathodes reported recently[40], implying that further innovative strategies are required.

Each dopant exerts a distinct influence on the properties of the host material, and achieving a synergistic compromise through the combination of different dopants might represent an effective way to tailor material properties more precisely[41, 42].

Indeed, the multi-dopant strategy has been widely employed within the SOFC field to enhance material characteristics, often leading to superior performance compared to oxides modified with a single dopant[43-45]. Therefore, it is reasonable to hypothesize that the performance of SFO-based cathodes for H-SOFCs can be significantly improved by adopting a multi-element doping strategy. In this study, the dopants Mo, Sn, Sc, and Zr were incorporated in equal molar ratios into the $Sr_2Fe_2O_6$ oxide, forming the multi-element doped $Sr_2Fe_{1.5}Mo_{0.125}Sn_{0.125}Sc_{0.125}Zr_{0.125}O_6$ (SFO-ZSSM) oxide. The suitability of M-SFO as a cathode for H-SOFCs was systematically evaluated and compared with SFO oxides tailored with single dopants (Mo, Sn, Sc, or Zr), thereby revealing the distinct advantages offered by the multi-element doping approach.

## 2. Experimental

The $Sr_2Fe_{1.5}Mo_{0.125}Sn_{0.125}Sc_{0.125}Zr_{0.125}O_6$ (SFO-ZSSM) oxide powder was synthesized via a conventional solid-state reaction method. Stoichiometric amounts of the precursor powders, $SrCO_3$, $Fe_2O_3$, $MoO_3$, $SnO_2$, $Sc_2O_3$, and $ZrO_2$, were mechanically mixed thoroughly using ball milling. The resulting mixture was then collected and calcined (fired) at 1400 °C for 5 hours in air to complete the solid-state reaction. X-ray diffraction (XRD) analysis was employed to examine the phase purity and crystal structure of the synthesized SFO-ZSSM powder. Scanning transmission electron microscopy coupled with energy-dispersive X-ray spectroscopy (STEM-EDS) was further employed to examine the elemental distribution and confirm the homogeneous incorporation of all dopants. For comparative purposes, a series of SFO-based oxides doped with single elements were also prepared using the identical solid-state reaction method. These single-doped materials included $Sr_2Fe_{1.5}Mo_{0.5}O_6$ (SFO-Mo), $Sr_2Fe_{1.5}Sn_{0.5}O_6$ (SFO-Sn), $Sr_2Fe_{1.5}Sc_{0.5}O_6$ (SFO-Sc), and $Sr_2Fe_{1.5}Zr_{0.5}O_6$ (SFO-Zr). The phase composition and crystal structure of all synthesized oxides (single-doped and multi-doped) were examined using XRD. The particle morphologies and microstructural features of the synthesized powders were observed using scanning electron microscopy (SEM).

To investigate the oxygen diffusion and surface exchange capabilities of these oxide materials, electrical conductivity relaxation (ECR) measurements were performed. Dense rectangular bars of SFO-Zr, SFO-Sc, SFO-Sn, SFO-Mo, and SFO-ZSSM were prepared by sintering pressed powder compacts. The DC four-point probe method was used to continuously monitor the electrical conductivity of each bar sample. During the ECR measurement, the testing atmosphere surrounding the sample was abruptly switched from dry air to a gas mixture containing 50% $O_2$ (balance inert gas). This step change in oxygen partial pressure induces a change in the oxide's conductivity as it approaches a new equilibrium state. The relaxation time required to reach this new conductivity equilibrium was recorded. This relaxation time is directly related to and reflects the kinetics of oxygen bulk diffusion and surface exchange processes in the oxide material. Similarly, the proton diffusion and surface exchange abilities can be evaluated using an analogous ECR method. In this case, the testing atmosphere is abruptly changed from dry air to humidified air. The incorporation of water vapor ($H_2O$) into the oxide lattice can lead to the formation of protonic defects, which subsequently alters the electrical conductivity of the material. The relaxation kinetics under this hydration/dehydration cycle provide insights into the proton transport properties.

To evaluate the electrochemical performance of the SFO-ZSSM oxide as a potential cathode for H-SOFCs, complete fuel cells were fabricated. The cells were based on a $BaCe_{0.7}Zr_{0.1}Y_{0.2}O_{3-\delta}$ (BCZY) proton-conducting electrolyte. The anode-supported half-cells, consisting of a NiO+BCZY anode substrate and a thin, dense BCZY electrolyte layer, were fabricated via a co-pressing and co-sintering technique; detailed preparation procedures for these NiO+BCZY/BCZY half-cells can be found in prior literature. Subsequently, a cathode slurry containing the SFO-ZSSM powder was deposited onto the surface of the sintered BCZY electrolyte. This assembly was then co-fired at 900 °C for 2 hours to establish good adhesion and interfacial contact, thus forming a complete fuel cell. For direct comparison, identical cells were fabricated using SFO-Zr, SFO-Sc, SFO-Sn, and SFO-Mo as the cathode materials, respectively, following the same deposition and firing procedure. The electrochemical performance of all fabricated cells was tested using humidified hydrogen (wet $H_2$) as the fuel and static air as the oxidant. Current-voltage (I-V) curves and corresponding power density curves were recorded using a standard electrochemical workstation. The resistances of the cells under open-circuit conditions were analyzed in detail using electrochemical impedance spectroscopy (EIS).

## 3. Results and Discussions

XRD characterization was performed on the series of SrFeO$_3$-based cathode materials, including SFO-Zr, SFO-Sc, SFO-Sn, SFO-Mo, and the multi-doped SFO-ZSSM. The XRD patterns presented in Figure 1(a) reveal that all synthesized samples exhibit characteristic diffraction peaks corresponding to a typical perovskite crystal structure. Crucially, no detectable impurity peaks are observed, indicating that all dopant elements were successfully incorporated into the host lattice without forming secondary phases. To obtain precise crystallographic parameters, Rietveld refinement of the XRD data, as shown in Figure 1(b-f), was conducted using the GSAS software package, assuming the Pm-3m space group and a cubic structure model. The resultant goodness-of-fit values ($\chi^2$) were 0.81, 0.89, 0.99, 0.85, and 1.07 for SFO-Zr, SFO-Sc, SFO-Sn, SFO-Mo, and SFO-ZSSM, respectively, confirming the reliability and quality of the refinement results. Based on these refinements, the lattice parameters of the samples were determined to be 3.96 Å, 4.03 Å, 3.87 Å, 3.92 Å, and 4.00 Å, respectively.

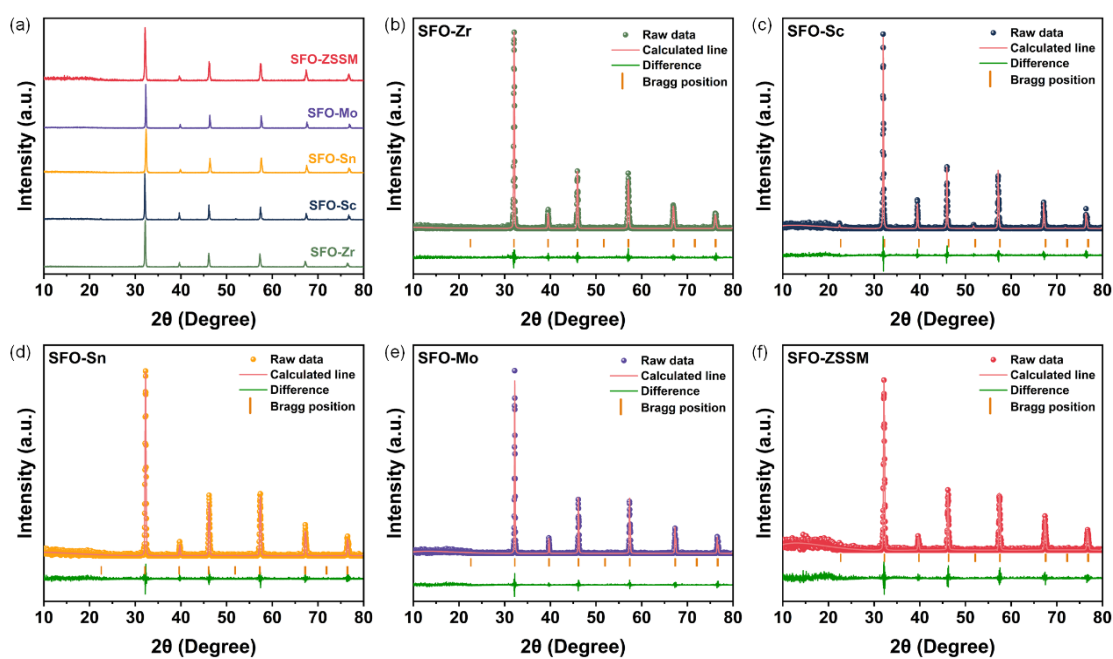

*Figure 1. (a) XRD patterns for as-prepared SFO-Zr, SFO-Sc, SFO-Sn, SFO-Mo and SFO-ZSSM. Rietveld refinement for (b) SFO-Zr, (c) SFO-Sc, (d) SFO-Sn, (e) SFO-Mo and (f) SFO-ZSSM.*

For a more detailed investigation into the phase structure of the multi-doped material, transmission electron microscopy (TEM) and high-resolution TEM (HR-TEM) characterizations were carried out specifically on the SFO-ZSSM sample. Elemental mapping analysis via STEM-EDS as shown in Figure 2(a) demonstrated a homogeneous distribution of all constituent dopants (Mo, Sn, Sc, Zr) and the host cations (Sr, Fe, O) without any noticeable segregation or clustering, providing direct verification of the successful and uniform incorporation of the four different elements into the perovskite lattice. As shown in Figure 2(b), by analyzing the spatial arrangement and angular relationships of the diffraction spots in the selected area electron diffraction (SAED) pattern, key crystal planes of SFO-ZSSM, namely the (1-10), (0-20), and (-1-10) planes, were identified. The corresponding interplanar spacings (d-spacings) were measured as 2.82 Å, 2.01 Å, and 2.82 Å, respectively. The lattice parameter calculated from these HR-TEM data was approximately 3.99 Å, which is in excellent agreement with the value of 4.00 Å obtained from the XRD Rietveld refinement. This consistency further confirms the high phase purity and structural integrity of the synthesized SFO-ZSSM oxide.

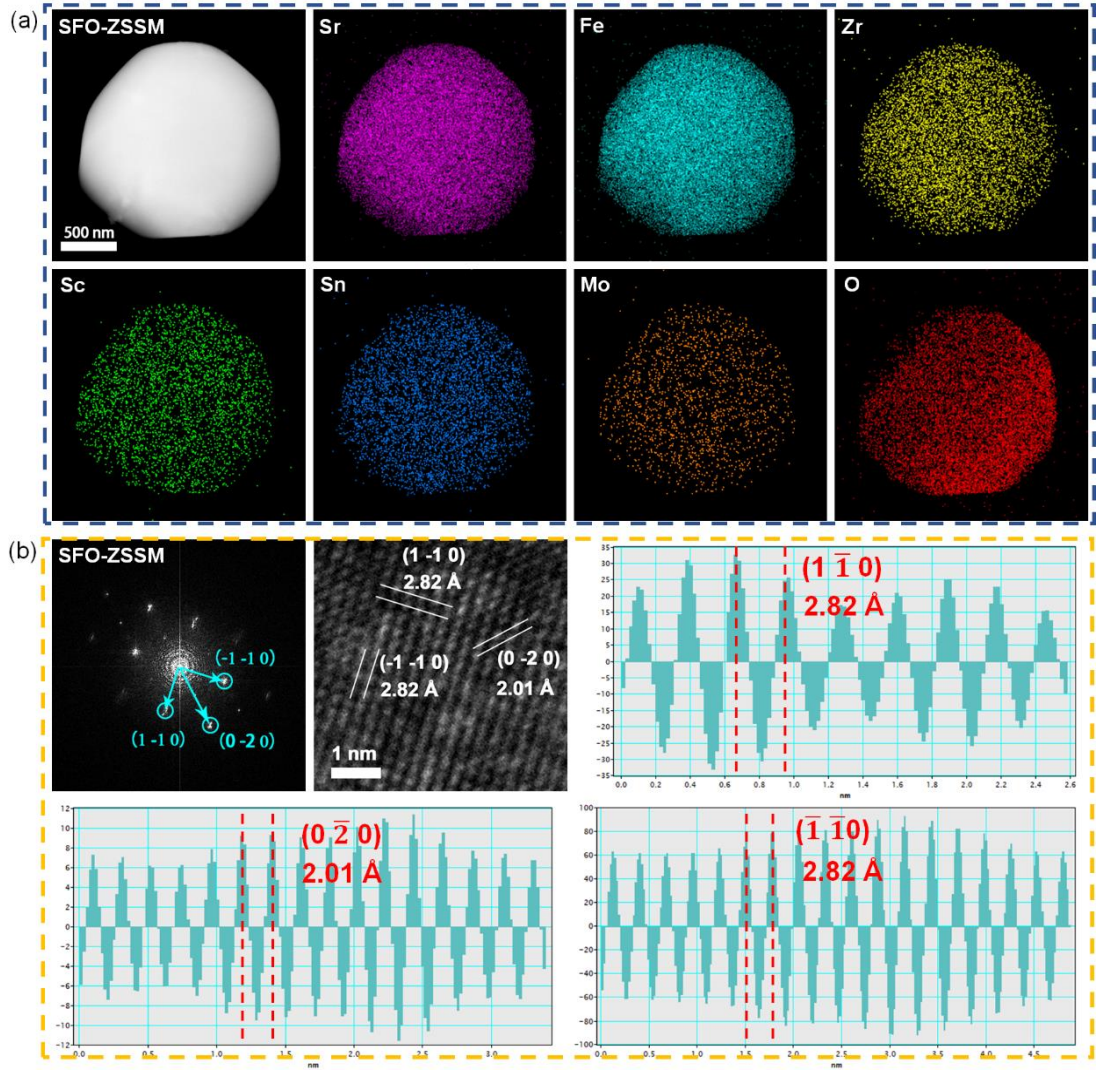

*Figure 2. (a) Elemental mapping and (b) SAED and HRTEM for SFO-ZSSM.*

Figure 3 presents the morphologies of the as-prepared SFO-Zr, SFO-Sc, SFO-Sn, SFO-Mo, and SFO-ZSSM powders as observed by SEM. It can be observed that all materials exhibit broadly similar microstructural features, indicating that the use of different single dopants does not significantly alter the overall powder morphology. Importantly, the adoption of the multi-element doping strategy in SFO-ZSSM also does not lead to a drastic change in microstructure compared to the singly-doped counterparts. Furthermore, although minor deviations exist, all materials possess similar average grain sizes. This observation implies that microstructural variations are unlikely to be the primary reason for the significant differences in electrochemical performance observed among these oxides.

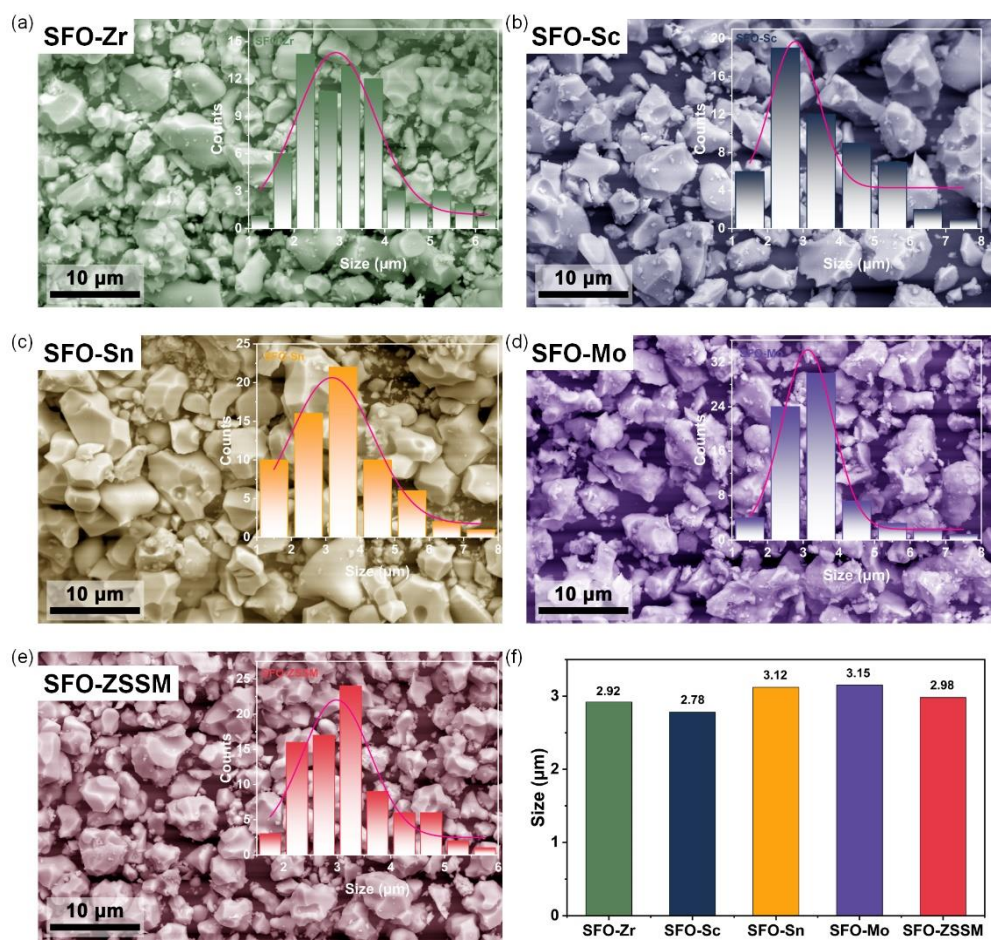

Figure 3. SEM image of the (a) SFO-Zr, (b) SFO-Sc, (c) SFO-Sn, (d) SFO-Mo, and (e) SFO-ZSSM powders with their particle size distribution. (f) The average size of these powders.

To study the influence of dopant selection on the oxygen and proton transport properties of the material, ECR measurements were performed on dense bar specimens of each oxide. Figure 4(a-e) shows the normalized conductivity relaxation curves for SFO-Zr, SFO-Sc, SFO-Sn, SFO-Mo, and SFO-ZSSM obtained under the condition of an abrupt atmosphere change from air to 50% $O_2$. It is evident that the relaxation time required to reach equilibrium for the singly-doped oxides (SFO-Zr, SFO-Sc, SFO-Sn,

SFO-Mo) exceeds 1000 seconds. In striking contrast, the relaxation time for the multi-element doped SFO-ZSSM is only a few hundred seconds. This marked reduction indicates that both oxygen diffusion and surface exchange kinetics are significantly accelerated by employing the multi-elemental doping strategy compared to any of the singly-doped oxides. By applying appropriate fitting models to the ECR curves, the chemical oxygen diffusion coefficient ($D_o$) and the surface exchange coefficient ($K_o$) can be quantitatively calculated; the results are summarized in Figure 4(f). Analysis reveals that different single dopants have distinctly different impacts on $D_o$ and $K_o$. Among the singly-doped materials, Zr appears to promote oxygen diffusion and surface exchange the most effectively, whereas Sn seems less optimal as it results in lower $D_o$ and $K_o$ values compared to the other samples. Notably, the $D_o$ and $K_o$ values of the multi-element doped SFO-ZSSM are not merely an average of the values from the four single-doped oxides. Instead, SFO-ZSSM exhibits the largest $D_o$ and $K_o$ values among all tested materials, demonstrating a clear synergistic effect. This synergy strongly indicates a distinct advantage of the multi-elemental doping strategy in promoting the oxygen diffusion and surface exchange abilities of SFO-based cathode materials.

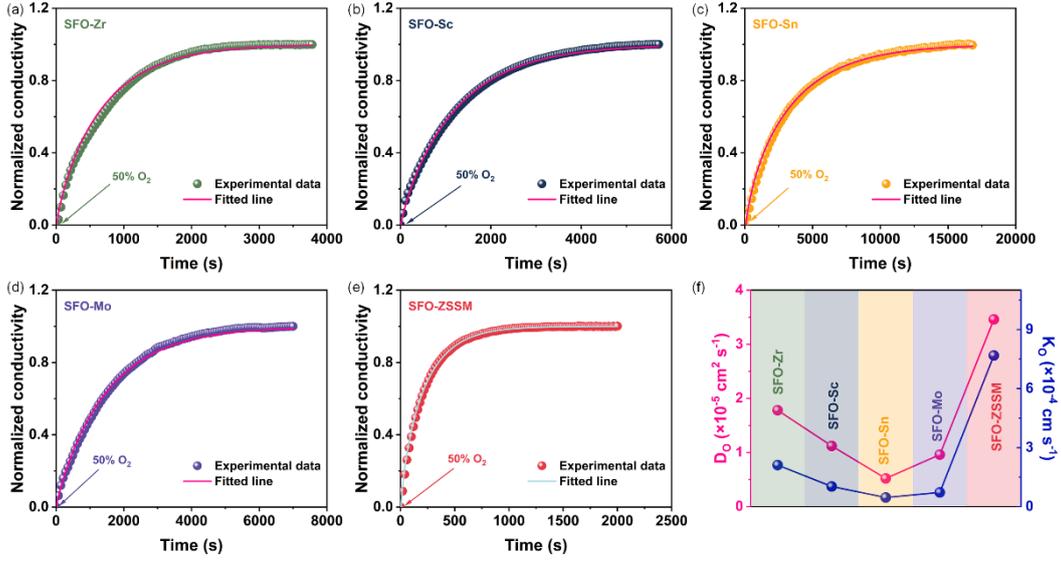

*Figure 4. ECR curves for a) SFO-Zr, (b) SFO-Sc, (c) SFO-Sn, (d) SFO-Mo, and (e) SFO-ZSSM in the condition of change the oxygen partial pressure. (f) Comparison of $D_o$ and $K_o$ for these oxides.*

Similar conclusions are drawn from measurements of the proton transport properties. The proton diffusion coefficient ($D_H$) and the corresponding surface exchange coefficient ($K_H$) were also determined via ECR. For a cathode operating in H-SOFCs, high proton diffusion ability is highly desirable because this feature can extend the electrochemically active reaction zone from the traditional triple-phase boundary to a broader two-dimensional area across the entire cathode surface[46-53]. Therefore, ECR measurements were used to investigate the proton diffusion and surface exchange abilities for the SFO-based oxides with different doping schemes. Figure 5(a-e) shows the ECR curves obtained when the atmosphere is abruptly changed from dry air to humidified air. The multi-elementally doped SFO-ZSSM consistently shows the shortest relaxation time among all tested samples. Regarding the singly-doped materials, a different trend is observed compared to the oxygen transport results. While Zr was most beneficial for $D_o$ and $K_o$, the Sc-doped sample (SFO-Sc) proves superior for proton transport, exhibiting larger $D_H$ and $K_H$ values than the other singly-doped oxides, as shown in Figure 5(f). This result clearly indicates that the influences of specific dopants on oxygen transport properties ($D_o$, $K_o$) and proton transport properties ($D_H$, $K_H$) are different and not necessarily correlated. Nevertheless, the multi-element doped SFO-ZSSM possesses the largest $D_H$ and $K_H$ values among all tested oxides, underscoring the advantage of the multi-element doping strategy in simultaneously enhancing the bulk diffusion and surface exchange abilities for protons.

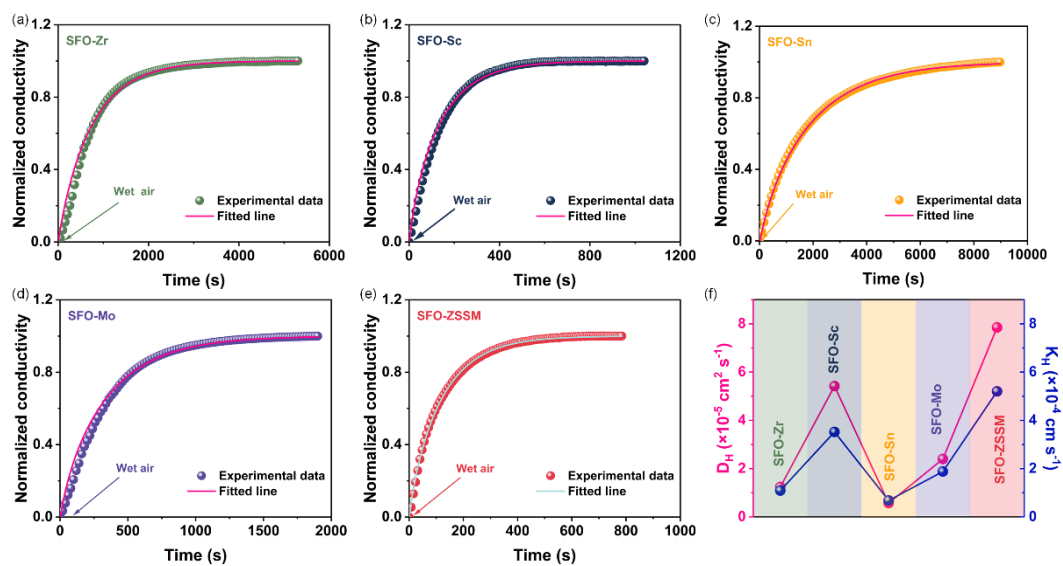

*Figure 5. ECR curves for (a) SFO-Zr, (b) SFO-Sc, (c) SFO-Sn, (d) SFO-Mo, and (e) SFO-ZSSM in the condition of change the water partial pressure. (f) Comparison of $D_H$ and $K_H$ for these oxides.*

Due to the concurrently improved oxygen and proton transport properties achieved via the multi-element doping strategy, it is expected that a fuel cell employing the SFO-ZSSM cathode would exhibit superior performance compared to cells using the singly-doped cathodes. Figure 6(a-e) shows the electrochemical performance of H-SOFCs utilizing SFO-Zr, SFO-Sc, SFO-Sn, SFO-Mo, and SFO-ZSSM cathodes, respectively. All cells show evident differences in their power output. The peak power densities (PPD) for the cell with the SFO-Zr cathode are 806, 509, and 388 mW cm$^{-2}$ at 700, 650, and 600 °C, respectively. This performance is significantly lower than that of the SFO-Sc cell, which achieves PPDs of 1058, 828, and 475 mW cm$^{-2}$ at the same respective temperatures. Although SFO-Zr exhibits the largest $D_o$ and $K_o$ among the singly-doped oxides, its positive impact on overall fuel cell performance is less profound than that of SFO-Sc, which boasts the highest $D_H$ and $K_H$. This comparison implies that, for H-SOFC cathodes, enhancement in proton diffusion and surface exchange kinetics has a more profound and direct influence on cell performance than improvement in oxygen transport properties alone. The cell with the SFO-Sn cathode shows the lowest PPDs (685, 387, 296 mW cm$^{-2}$) among all singly-doped oxides, which is consistent with its poorest performance in both oxygen and proton transport metrics (lowest $D_o$, $K_o$, $D_H$, $K_H$). In marked contrast, the multi-element doped SFO-ZSSM cathode, possessing the largest values for all four key transport parameters ($D_o$, $K_o$, $D_H$, $K_H$) among all tested oxides, also delivers the highest electrochemical performance. It achieves outstanding PPDs of 1580, 1137, and 854 mW cm$^{-2}$ at 700, 650, and 600 °C, respectively. As graphically summarized in Figure 6(f), the performance of the SFO-ZSSM-based cell is evidently superior to that of cells using any singly-doped cathode across the entire tested temperature range. The dramatically enhanced fuel cell output is primarily attributed to the significantly improved proton diffusion and surface exchange abilities, which greatly extend the active reaction area within the cathode. Additionally, the concurrently increased $D_o$ and $K_o$ values further enhance oxygen reduction kinetics, contributing synergistically to the overall improvement in cathode activity.

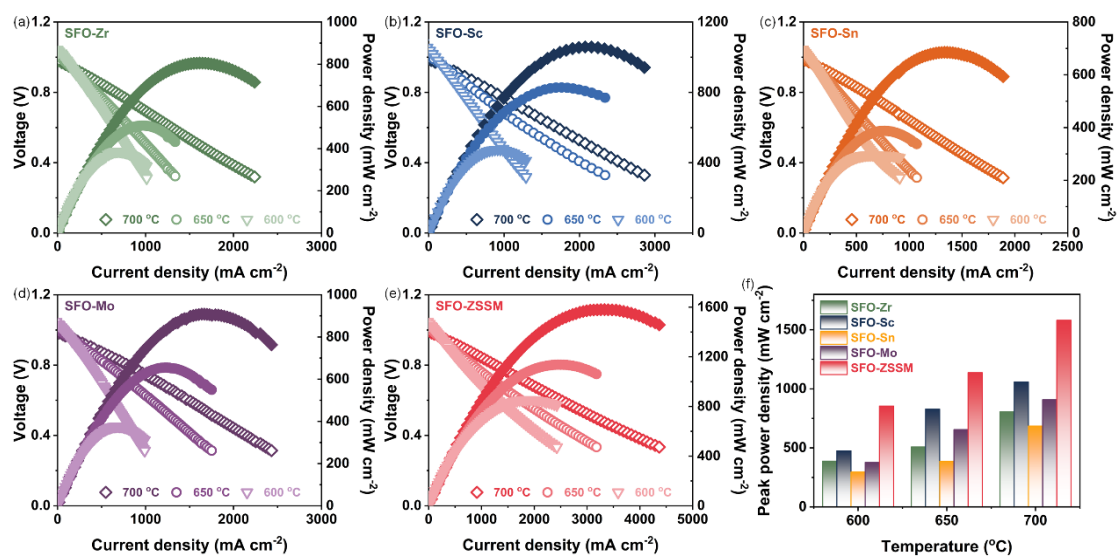

*Figure 6. Cell performance of fuel cells using (a) SFO-Zr, (b) SFO-Sc, (c) SFO-Sn, (d) SFO-Mo, and (e) SFO-ZSSM cathodes. (f) PPD comparison for these cells.*

Figure 7(a-e) displays the EIS plots, measured under open-circuit conditions at 700 °C, for cells employing the different cathodes. All cells exhibit a similar ohmic resistance ($R_o$), which primarily comprises contributions from the ionic resistance of the BCZY electrolyte and the interfacial contact resistances. This result is expected since all cells were fabricated using identical half-cells (same electrolyte material and thickness) and sintering conditions. In contrast to the consistent $R_o$, these cells show significantly different polarization resistances ($R_p$). The $R_p$ values for cells with SFO-Zr, SFO-Sc, SFO-Sn, SFO-Mo, and SFO-ZSSM cathodes are 0.098, 0.047, 0.121, 0.068, and 0.034 $\Omega$ cm$^2$, respectively. It is clear that the $R_p$ of the SFO-ZSSM cell is substantially smaller than that of any cell using a singly-doped cathode. Given that all cells share the same anode, the observed differences in $R_p$ should originate primarily

from the cathode side, strongly suggesting that the superior catalytic activity of SFO-ZSSM is the main reason for its reduced polarization resistance. As summarized in Figure 7(f), while all cells show comparable $R_o$, their $R_p$ values differ markedly, leading to different total cell resistances ($R_{tot} = R_o + R_p$). Owing to its smallest $R_p$, the SFO-ZSSM cell possesses the smallest $R_{tot}$, which directly correlates with its highest power output. Figure 7(g) presents cross-sectional SEM images of the tested cells with different cathodes. All cells exhibit similar overall morphologies without major microstructural differences such as delamination or abnormal grain growth. Furthermore, the cathode layer adheres well to the electrolyte surface in all cases. These observations collectively indicate that the pronounced differences in electrochemical performance among the cells are primarily attributable to the intrinsic material properties (i.e., catalytic and transport properties) of the cathodes, rather than to variations in cell microstructure.

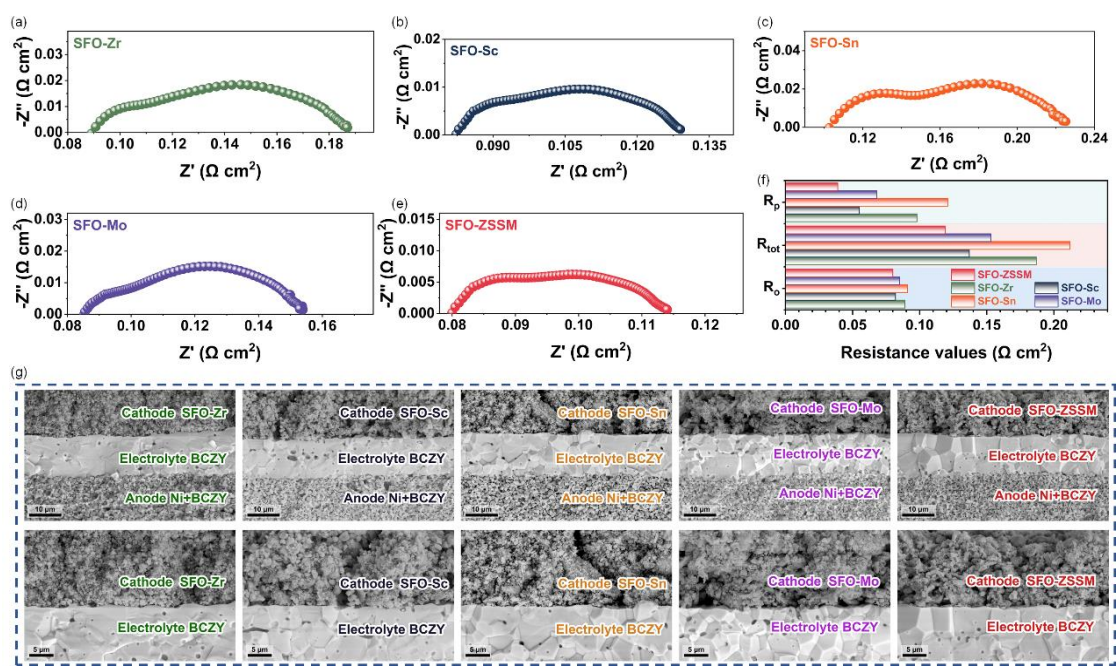

*Figure 7. EIS plots of fuel cells using (a) SFO-Zr, (b) SFO-Sc, (c) SFO-Sn, (d) SFO-Mo, (e)*

*SFO-ZSSM cathodes, and (f) their comparison in $R_o$, $R_p$ and $R_{tot}$. (g) SEM of the tested cells.*

In addition to its excellent electrochemical performance, the fuel cell utilizing the SFO-ZSSM cathode demonstrates promising operational stability under typical working conditions. Figure 8(a) shows the cell voltage as a function of time under a constant applied current density. The voltage remains remarkably stable for almost 100 hours of continuous operation, with no evident degradation detected, suggesting good short-term operational stability for the SFO-ZSSM cathode. Figure 8(b) and (c) further compares the I-V curves and EIS plots of the SFO-ZSSM cell before and after the long-term stability test. The peak power density remains almost unchanged after the extended operation, and no significant increase in total cell resistance is observed, which provides further evidence for the good stability of the cell. The robust stability of the cell may be partially attributed to the chemical composition of SFO-ZSSM. Unlike many Ba-containing proton-conducting ceramics, this material is barium-free, which should inherently confer better chemical stability against common atmospheric contaminants like $CO_2$ and moisture ($H_2O$). Furthermore, post-test microstructure observation of the cell, as shown in Figure 8(d) and (e), indicates that no degradation or destruction of the cathode or interface microstructure can be observed, and the cathode maintains excellent adhesion to the electrolyte. In addition, no evident cathode and electrolyte interfacial reaction can be detected. This preserved structural integrity constitutes another important reason for the observed stable fuel cell performance[47].

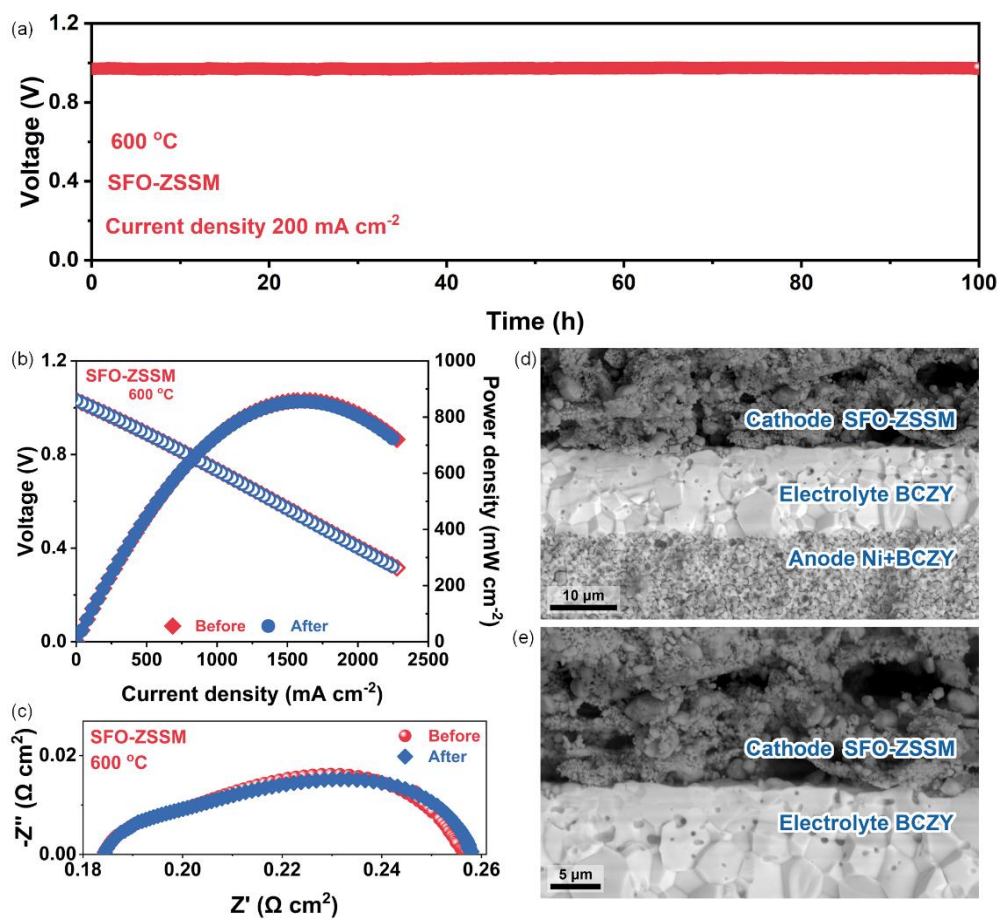

*Figure 8. (a) Long-term stability of the cell using the SFO-ZSSM and the (b) IV and (c) EIS plot comparison of the cell before and after the long-term test. SEM image of (d) the complete cell and (e) cathode/electrolyte interface.*

## 4. Conclusions

In this work, a novel multi-element doped SFO-based cathode, SFO-ZSSM, was proposed to address the performance limitations of cobalt-free cathodes for H-SOFCs. The central hypothesis—that a synergistic combination of dopants could more effectively tailor the functional properties than single-dopant modifications—was rigorously tested. Structural analyses confirmed that all dopants were successfully incorporated into a single-phase perovskite lattice without segregation. The key finding from ECR studies was that the multi-element doping strategy induced a remarkable synergistic enhancement in both oxygen and proton transport properties, far exceeding the performance of any singly-doped oxide. This superior oxygen and proton diffusion and surface exchange capabilities directly translated to outstanding electrochemical performance. When deployed in complete fuel cells with a BCZY electrolyte, the SFO-ZSSM cathode delivered peak power densities substantially higher than those achieved with any single-doped SFO cathodes. Impedance analysis confirmed that this performance enhancement originated from a significantly reduced cathode polarization resistance, a direct consequence of its accelerated surface exchange and bulk diffusion processes. Notably, the results highlighted that for H-SOFC cathodes, the enhancement of proton transport kinetics had a more profound impact on the overall cell performance than the improvement of oxygen transport alone. Furthermore, the SFO-ZSSM-based cell demonstrated excellent short-term stability under operating conditions, with no significant performance degradation over 100 h. This stability was attributed to the preserved cathode-electrolyte interface adhesion and the inherent chemical stability of the Ba-free composition. This study successfully demonstrated that the multi-element doping strategy is a powerful and effective method for engineering high-performance, cobalt-free cathode materials for H-SOFCs. The synergistic effects achieved in SFO-ZSSM led to unparalleled ionic conductivity and catalytic activity, establishing a new promising candidate and a viable material design principle for advancing H-SOFCs.

## Acknowledgements

This work was supported by the National Natural Science Foundation of China (Grant Number: 52302263 and 52272216).